\documentclass[epsf,preprint,aps]{revtex4}
\usepackage{graphicx}
\usepackage{bm}

\date{\today}

\begin{document}

\title{CRYSTAL Simulation Code and New Coherent Effects in Bent Crystal at the LHC}

\author{A.~I.~Sytov, V.~V.~Tikhomirov}
\email{alex_sytov@mail.ru, vvtikh@mail.ru} \affiliation {Research Institute for Nuclear Problems, Bobruiskaya str., 11, 220030, Minsk, Belarus.}

\begin{abstract}
The LHC crystal-based collimation system is mainly addressed. A CRYSTAL simulation code for particle tracking in crystals is introduced. Its essence consists in both adequate and fast sampling of proton trajectories in crystals which is crucial for both correct description of experiments and quantitative prediction of new effects. The H8 single-pass experiment at the CERN SPS as well as 7 TeV proton deflection by a bent crystal at the LHC are simulated. We predict the existence of dechanneling peaks corresponding to the planar channeling oscillations as well as describe the possibility of their observation at high energies, specifically in the LHC crystal-assisted collimation experiment planned on 2015. An effect of excess over the amorphous level of ionization losses in the channeling mode was also found for the LHC energy. In addition, the LHC crystal-based collimation system is simulated as well as its possible improved layouts with application of a crystal with the cut and multiple volume reflection in one bent crystal.
\end{abstract}
\maketitle

\section{Introduction}

The problem of the LHC collimation efficiency is very important for the next collider runs. Since the LHC beam intensity is constantly rising, the radiation load on the sensible accelerator equipment like superconducting magnets will also increase. The problem of the magnet quench \cite{quench} represents one of the main threats for stable functioning of the collider. The crystal-based collimation is expected to be a simple and smart solution allowing one to considerably decrease the level of particle losses in the most critical parts of the accelerator. In addition, it will possibly reduce a parasitic background in the LHC detectors as it was achieved at Tevatron \cite{Tevatron}.

The LHC crystal-based collimation experiment planned on 2015 \cite{LUA9} is dedicated to the application of the channeling effect in a bent crystal for controlled removal of a beam halo and its subsequent absorbtion in secondary collimators. In the simplest case the corresponding collimation layout consists of a bent crystal as a primary deflector and an absorber.

The applicability of the channeling mode strongly depends on the angular divergence of an initial beam. If it is several times smaller than the critical channeling angle (Lindhard angle) as it is envisaged for the LHC \cite{LUA9} the single-pass channeling efficiency may exceed 80-85\% \cite{CrysColl}. Otherwise the efficiency of a crystalline deflector will be strongly reduced.

For both of these cases smart modifications were proposed. For very small angular divergence a crystal with the cut is capable to increase the channeling single-pass efficiency up to 98-99\% \cite{cut,amlayer}. This application is potentially useful for possible projects of beam extraction with a bent crystal from the LHC \cite{LHCExtr} or 100 TeV collider \cite{FCCExtr}. But if the divergence is too large, the effect of multiple volume reflection in one bent crystal \cite{MVROC,MVROCTGM} and its modifications \cite{MVROCTS} becomes the most efficient for the LHC.

Correct simulations of particle trajectories in a crystal is crucial for the adequate treatment of experiments. We argue that simulations without the evaluation of trajectories are not sufficient for the adequate experiment description. In addition the prediction of new effects in simulations is possible only for a realistic numerical experiment but not for adjustments of already known effects. In opposite, the latter even in a good coincidence with an experiment for one energy may lead to complete discordance for the other one.

We introduce a CRYSTAL simulation code providing both correct and fast simulations of charged particles trajectories in bent crystals. A good agreement of  simulation results with the single-pass experiments at the SPS \cite{H8} confirms a high enough accuracy of our simulation tool. Additionally, the anticipation of the new effects of dechanneling peaks and an excess of the amorphous level of ionization losses in the channeling mode demonstrates the predictive capabilities of the program.

We apply our simulation technique to the planned LHC crystal-based collimation experiment \cite{LUA9} as well as to its possible configurations with application of a crystal with the cut \cite{cut,amlayer} and the effect of multiple volume reflection in one bent crystal \cite{MVROC,MVROCTGM,MVROCTS}.

\section{CRYSTAL simulation code and its application to the H8 single pass experiment at the CERN SPS}\label{s1}

The main conception of the CRYSTAL software is simulation of particle trajectories by solving of equations of motion. The CRYSTAL code includes both 1D and 2D models. A 2D model is used for calculation of particle motion in an axial potential $U(x,y)$ depending on transverse coordinates $x$ and $y$ and averaged along the longitudinal coordinate $z$. The planar potential $U(x)$ is obtained by averaging along the coordinate $y$. In both cases the potential is averaged over thermal vibrations. The relativistic Lorentz equation is solved numerically by the 4th order Runge-Cutta method 3/8 rule \cite{RungeCutta}. The integration step $\Delta z$ is chosen to be small enough for particle to pass for one step in a transverse direction no more than 1/500 of the interplanar or interaxial distance.

Another essential feature for the trajectory simulations is Coulomb and nuclear incoherent scattering. Coulomb scattering on both nuclei and electrons naturally divides on multiple and single processes \cite{Tikh1,Tikh2}. Multiple scattering is simulated cumulatively including single processes only at sufficiently small angles. Single Coulomb scattering is simulated in details with the application of Rutherford cross-section. Electron $n_{e}(x,y)$ density is obtained according to the Tomas-Fermi approximation of screened atomic potential.

The elastic and quasielastic (diffractive) scattering r.m.s. angles were calculated according to \cite{MokhovSc}, the Gaussian angular distribution was used. For the inelastic scattering case in a crystal a particle was considered to be lost.

For the main functions of the electric potential $U(x,y)$, its first derivatives, nuclear $n_{N}(x,y)$ and electron $n_{e}(x,y)$ densities spline interpolations were used. The interpolation coefficients are saved into input files to reproduce the function values during the program run. Such a technique possesses two advantages. The first one is universality of an algorithm making possible exchange of the potential without the program modifications. In our calculations we used either Moliere \cite{Biryukov} or Doyle-Terner \cite{DoyleTerner} potential. The second advantage is a high simulation speed provided by a small number of mathematical operations needed for spline interpolation. At the same time its accuracy is can be made quite high and checked before the program run. For the LHC case (see below) the CRYSTAL simulation speed is up to $\sim$ 2000 proton trajectories per second per CPU.

The H8 experiment \cite{H8} is dedicated to studying of beam deflection by a bent crystal. The main idea is to measure the difference between outcoming and incoming angles of 400 GeV protons.

The CRYSTAL simulation results are shown in Fig. 1 together with the experimental ones. A good agreement is demonstrated. The simulation statistics is 8 times higher than the experimental one. It is important to stress that the scattering on experimental equipment must be taken into account. Otherwise the obtained results would be considerably different.

\begin{figure}[b]
\hspace{-0.0cm} \vspace{0cm} {\includegraphics[width=10cm]{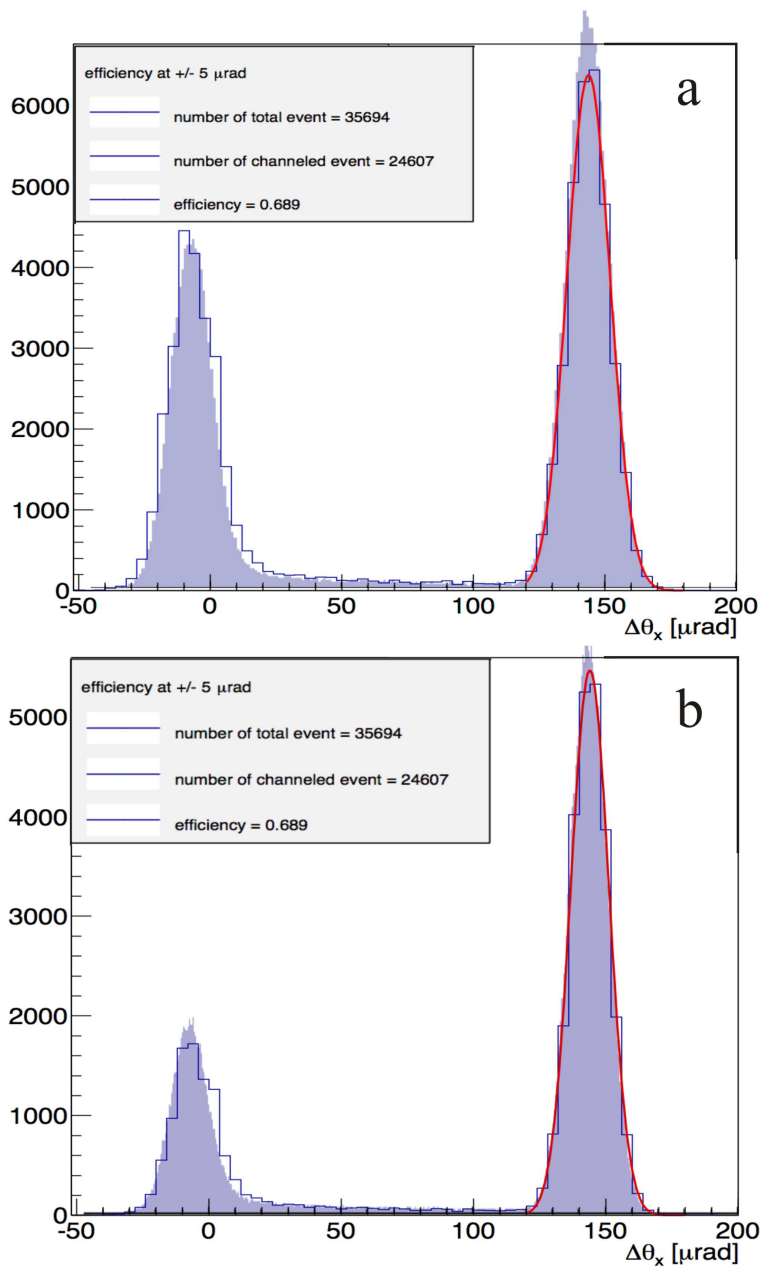}} \caption {The deflection angle distribution (the difference between incoming and outcoming angle) after the STF45 crystal passage in the H8 line at the CERN SPS. Only the intervals of measured incoming angles of $\pm$10 $\mu$rad (a) and $\pm$5 $\mu$rad (b) were selected.}\label{F1}
\end{figure}

The obtained (during the test without the crystal in the line) r.m.s. scattering angle on detectors achieved $\sim$ 6.3 $\mu$rad \cite{H8no}. This value should be divided between two detector arms  \cite{H8line} (for incoming and outcoming angle measurement). For equal separation the angle should be divided by $\sqrt{2}$ which will give the value of $\sim$ 4.5 $\mu$rad for each arm. This angle includes scattering on detector planes, air, etc. A Gaussian distribution of scattering was taken for both input and output angle. The measured initial angular distribution as well as the crystal parameters were chosen according to \cite{H8}.

The coincidence between experimental results and simulation was obtained from "the first principles" without any adjustment of the trajectory evaluation. This demonstrates the predictive power of the method using the direct trajectory simulations. One can conclude also that the crystal alignment accuracy in the experiment corresponds to the 1 $\mu$rad level required for the LHC.

\section{New effects for beam deflection at the LHC}\label{s3}

The LHC case is specific by very high energy and usually low number of channeling oscillations of about 4 per mm. Therefore, if the angular divergence is rather low, as for the distribution considered, the phase correlation between different trajectories will become observable. This situation is useful to underline specific features of the CRYSTAL routine.

\begin{figure}[b]
\hspace{-0.0cm} \vspace{0cm} {\includegraphics[width=10cm]{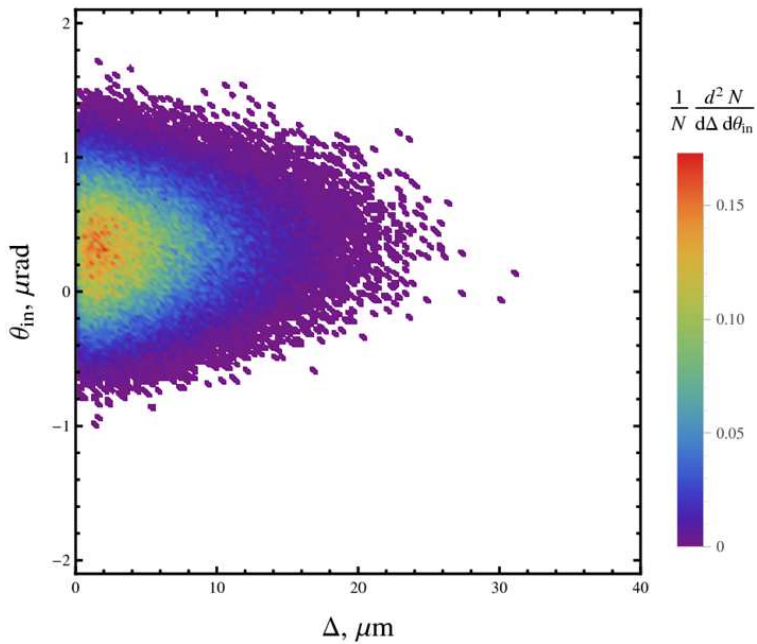}} \caption {The initial distribution of 7 TeV halo protons hitting the bent crystal in the planned crystal-based collimation experiment \cite{LUA9}.}\label{F2}
\end{figure}

The angular distribution behind the crystal at channeling orientation is shown in Fig. 3. The crystal parameters were taken the same as in the LHC crystal-based collimation system \cite{LUA9}. The initial beam distribution in coordinates and angles (Fig. 2) was calculated according to the model of beam slow diffusion \cite{miscut} considering both betatron and synchrotron oscillations. The diffusion step of betatron amplitude increment per turn was chosen to obtain the average impact parameter of 5 $\mu$m as in \cite{LUA9old}. This distribution is used in all the LHC simulations in this paper. The crystal alignment cannot be ideal because of the presence of synchrotron oscillations. That is why it was taken equal to 0.5 $\mu$rad.

A new feature is observed for dechanneling. Namely, in CRYSTAL simulations the \textit{dechanneling peaks} corresponding to the planar channeling oscillations \cite{Gemmel,Feldman,Ba71} are visible. Planar channeling oscillations being already observed for protons at non-relativistic energies \cite{PCO0,PCO1,PCO2} can be revealed for the relativistic ones.

It is important to stress that the dechanneling peaks visible even at zero beam angular divergence. Therefore, it could not be connected with initial distribution details. The main reason of their appearance consists in the preservation of the phase correlations of different trajectories at large distances. The dechanneling peaks appear when particles approach to the crystal planes where the probability of scattering on nuclei is high. The minimum of dechanneling probability corresponds to the maximal distance from the crystal planes where the nuclear density is minimal. Thus, the number of dechanneling peaks is simply to that of the particle approaches to the crystal planes. For the 4 mm crystal at 7 TeV this number is equal to 32. Approximately this value is observed in Fig. 3.

\begin{figure}[t]
\hspace{-0.0cm} \vspace{0cm} {\includegraphics[width=10cm]{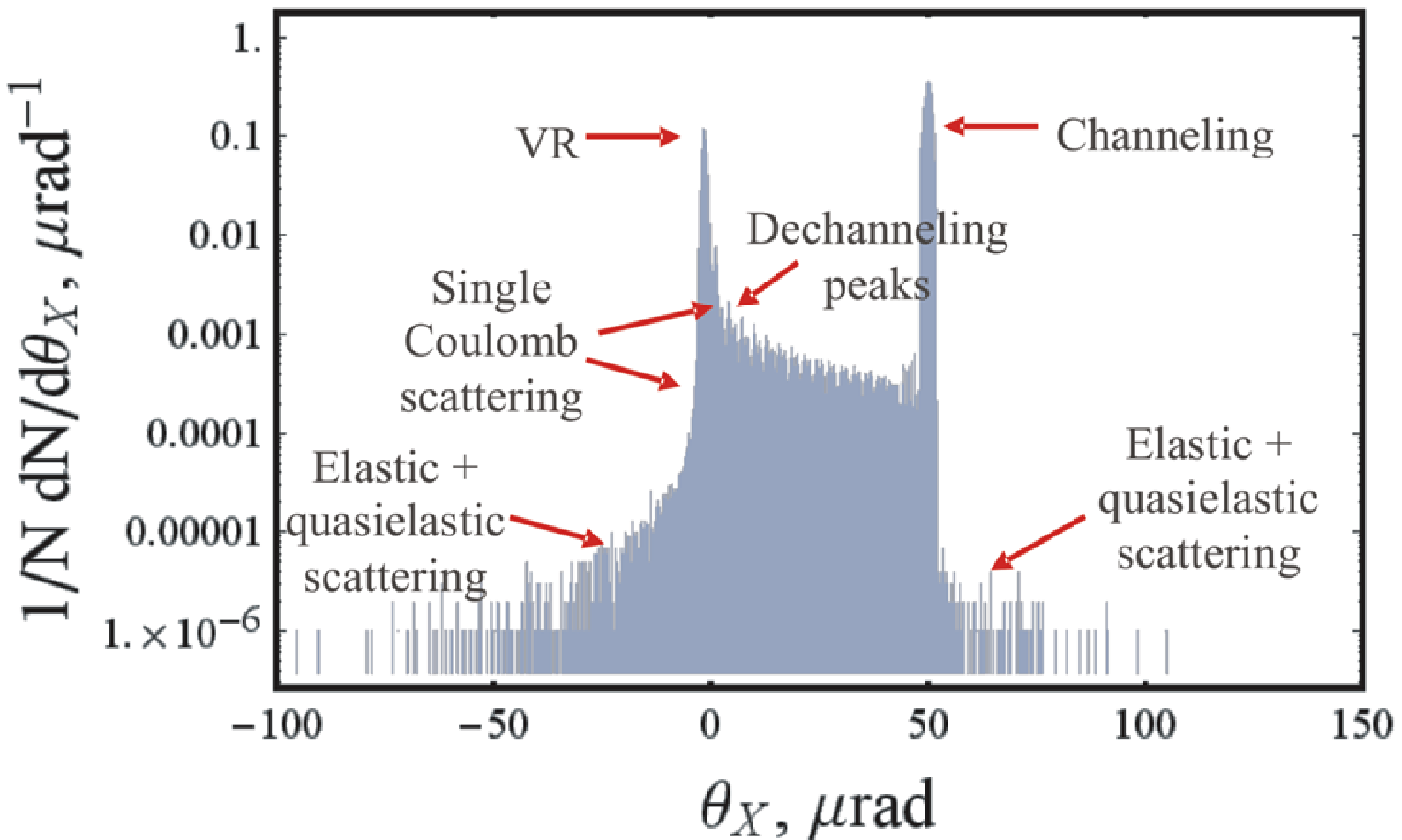}\\ \includegraphics[width=10cm]{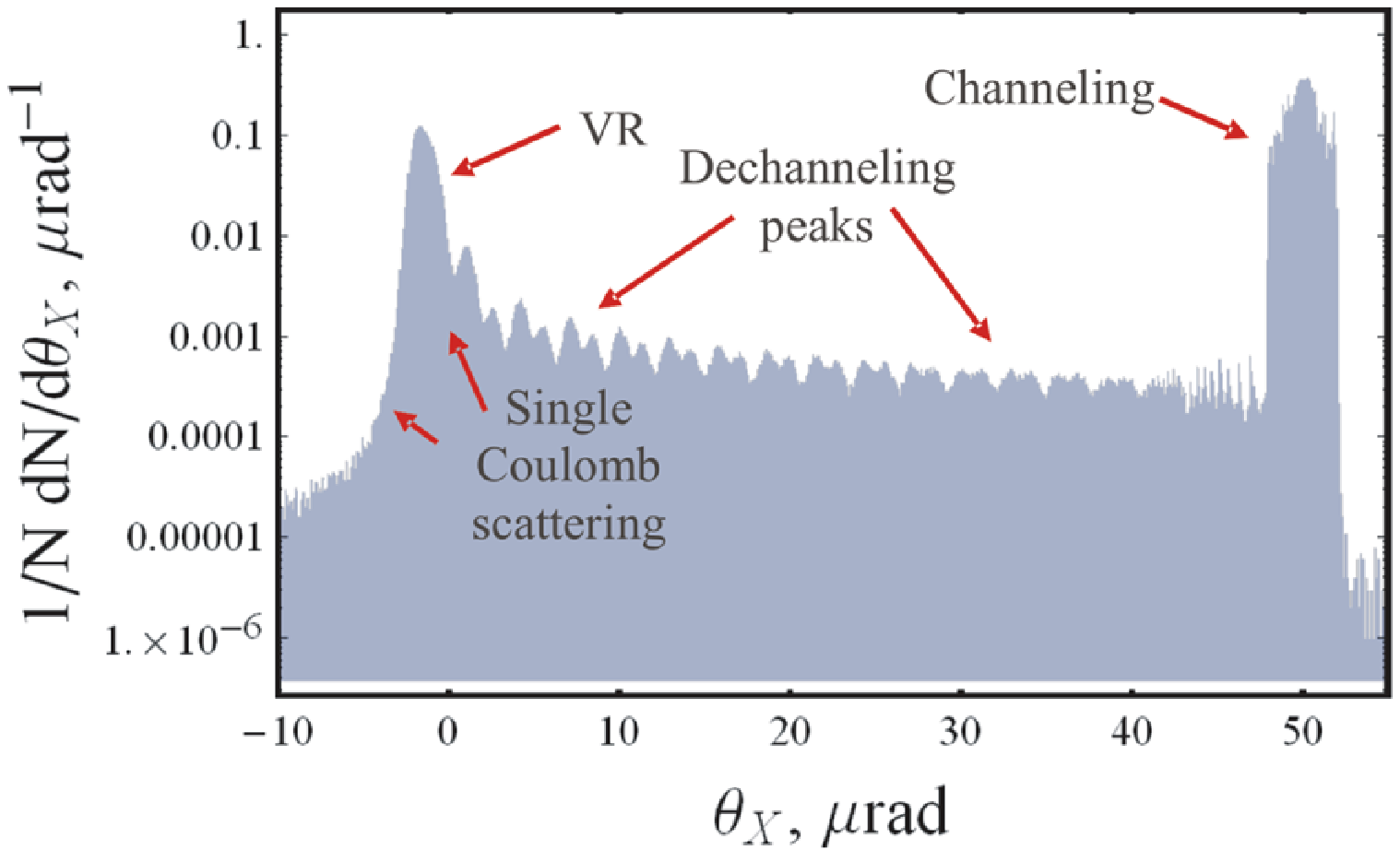}} \caption {The angular distribution of 7 TeV protons behind the bent Si (110) crystal of 4 mm length along the beam and 50 $\mu$rad bending angle at the channeling orientation.}\label{F3}
\end{figure}

To make the dechanneling peaks observable the Coulomb scattering angle in the crystal should be less than the half of the angular distance between the peaks:
\begin{equation}
\label{5}
\frac{\Delta \varphi _{peak}}{2 \theta _{sc}}=\frac{\lambda  \theta _b}{4 l_{cr}}\frac{pv}{13.6 MeV \sqrt{l_{cr}/X_r} \left(0.038 \ln \left(l_{cr}/X_r\right)+1\right)}>1,
\end{equation}
where $l_{cr}$ is the crystal length, $\theta _b$ is its bending angle, $\lambda$ is the channeling oscillations length defined by the potential well shape and the particle energy \cite{Biryukov}, $X_{r}$ is the radiation length equal for silicon to 9.36 cm \cite{PDG}, $\theta _{sc}$ is the r.m.s. Coulomb scattering angle in the crystal which was estimated according to \cite{PDG}. Putting $l_{cr}=4$ mm, $\theta _b=50$ $\mu$rad, $pv=7$ TeV as for the planned LHC crystal-based collimation experiment \cite{LUA9} we will obtain the estimate $\Delta \varphi _{peak}/2 \theta _{sc}\approx2$ demonstrating that the condition is indeed fulfilled.

Another important condition is rather small angular divergence of the initial beam -- at least 2-3 times smaller than the Lindhard angle \cite{Biryukov}. This will provide a good phase correlation of different trajectories. This condition is expected to be satisfied for the LHC \cite{LUA9}.

One can also connect the different height of these peaks with the asymmetry of the potential well (Fig. 4). Because of a strong centrifugal force particles will dechannel much more frequently at the low (left) potential maximum. Here it should be mentioned that 7 TeV protons dechannel very close to the crystal planes. That is the higher dechanneling peaks correspond to the moment of close approach to the left potential maximum. Moreover, the number of particles captured into the channeling mode in the dechanneling zone is higher at the left maximum of the potential than near the opposite particle reflection point since the dechanneling zone near the left maximum is wider than the zone of the same transverse energy near the right one (see Fig. 4). That is why the different height of dechanneling peaks is directly connected with the asymmetry of potential well.

\begin{figure}[b]
\hspace{-0.0cm} \vspace{0cm} {\includegraphics[width=10cm]{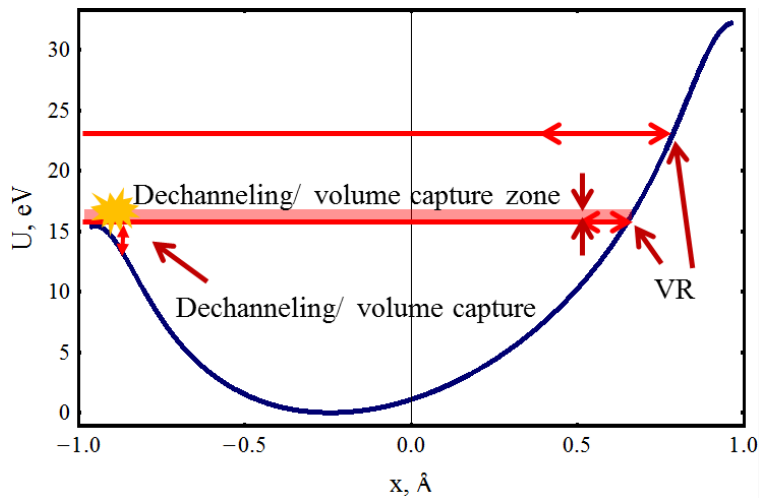}}
 \caption {The interplanar potential well for the bent Si (110) crystal of 4 mm length and 50 $\mu$rad bending angle.}\label{F4}
\end{figure}

Only high dechanneling peaks are visible for the volume reflection orientation (see Fig. 5). It can also be explained by a good phase correlation of transverse motion of volume captured particles, even better than for the channeling orientation. This phase correlation appears because a particle can be captured with a high probability only near the left potential peak due to the asymmetry of the potential well (see Fig. 4). That's why the transverse oscillation phases of different particles after the capture will be very close.

\begin{figure}[t]
\hspace{-0.0cm} \vspace{0cm} {\includegraphics[width=10cm]{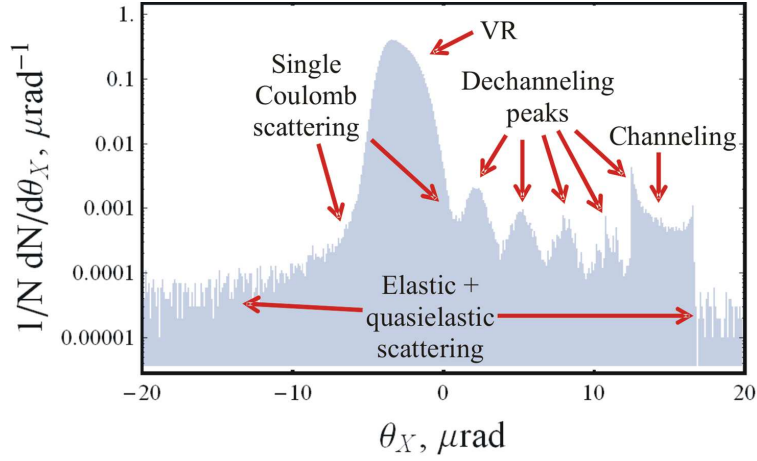}}
 \caption {The angular distribution of 7 TeV protons behind the (110) bent crystal of 4 mm length and 50 $\mu$rad bending angle and volume reflection orientation is $-35$ $\mu$rad.}\label{F5}
\end{figure}

It is important that all the effects mentioned above are really observable. The peak form naturally depends on the initial beam distribution. However, in the case of small enough angular divergence required for most particles to be captured in the channeling regime at 7 TeV all these effects can be revealed.

It is important to stress here that nothing forbids the dechanneling peaks observation at lower energies. The higher energies are more favorable for this simply because the averaged angle of nuclear scattering is inversely proportional to energy $\theta _{sc}\sim 1/pv$ while the Lindhard angle -- to its square root $\theta _{L}\sim 1/\sqrt{pv}$. This circumstance decreases the influence of scattering with the energy increase. That's why the dechanneling peaks should appear naturally at the LHC while at the SPS the crystal should be shorter and bent stronger than that in the previous section. For the STF45 crystal the ratio defined by Eq. (\ref{5}) is 0.24 i.e. considerably less than one. So, the peaks could not be observed in this case. However, it becomes possible for the crystal STF49 \cite{H8} ($l_{cr}=0.8$ mm, $\theta_{b}=247$ $\mu$rad) for rather small angular divergence of the initial beam (see Fig. 6, the ratio (\ref{5}) equals 1.7).

\begin{figure}[b]
\hspace{-0.0cm} \vspace{0cm} {\includegraphics[width=10cm]{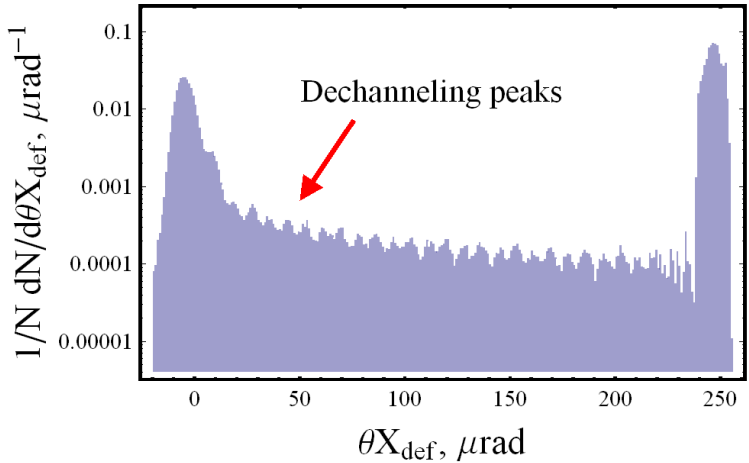}}
 \caption {The angular distribution of 400 GeV protons (SPS) behind the STF49 bent crystal of 0.8 mm length and 247 $\mu$rad bending angle; channeling orientation. R.m.s. beam angular divergence is equal to 2 $\mu$rad.}\label{F6}
\end{figure}

The ionization losses map (particles distribution on both deflection angles and energy losses) at the channeling orientation is shown in Fig. 7. The channeled particles with high amplitude of oscillations can lose even more energy than in the amorphous silicon. This is because the particles spend considerable time near crystal planes where the electron density is much higher than the electron density of the crystal. Thus, the electron density averaged along the trajectory will be higher than its average value in the crystal. These particles dechannel fast as a consequence.

The trajectory correlations are also noticeable. The ``waves'' at the dechanneling zone represent dechanneling peaks. The loop-like picture for the channeled particles is explained by the different number of channeling oscillations as well as by the different level of ionization losses depending on the oscillation amplitude.

\begin{figure}[t]
\hspace{-0.0cm} \vspace{0cm} {\includegraphics[width=10cm]{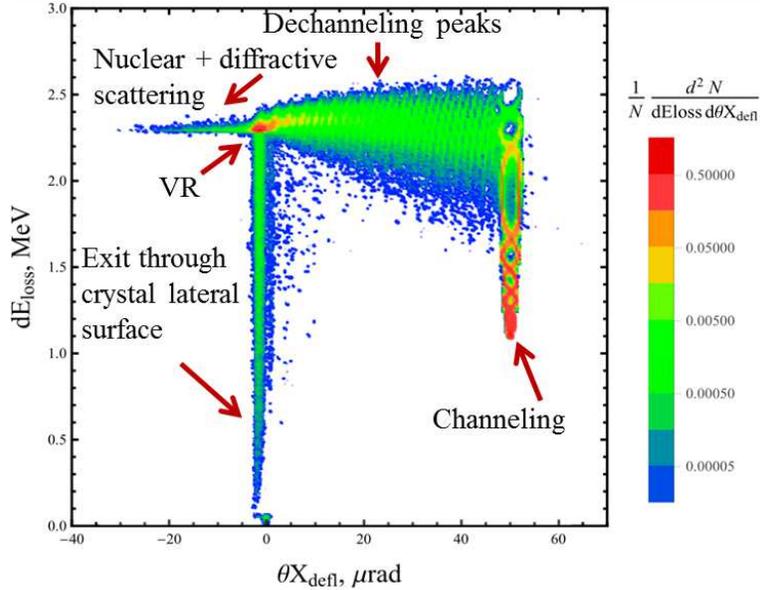}} \vspace{0cm}
 \caption {Ionization losses map at channeling orientation.}\label{F7}
\end{figure}

The results shown at the ionization losses map could be tested experimentally in the case of high enough resolution of the experimental equipment. The problem of correct simulation of ionization losses is related to the correct description of the electron density in crystal channels. As it was mentioned above, we describe the electron density in the frame of the Tomas-Fermi model with averaging over thermal vibrations \cite{Biryukov}. However, the electron density as well as the ionization loss model should be verified.

\section{Crystal-based collimation at the LHC}\label{s4}

The LHC crystal-based collimation system was designed to protect the superconducting magnets from halo particles. In this experiment at the LHC \cite{LUA9} the two crystals of 4 mm length, 50 mm height, 2 mm thickness and 50 $\mu$rad bending angle will be used to clean the halo of 7 TeV proton beams. The crystal deflecting the horizontal beam halo is bent along the (110) plane as well as another one for vertical halo is bent along the (111) plane. Both of them are placed in 19919.24 m point of the LHC at $6\sigma$ \cite{LUA9}. In this paper only the horizontal configuration of the collimation system was simulated.

The LHC-type collimators are double-sided, i.e. there are 2 collimators symmetrically placed from both sides of the beam. There are a lot of collimators at the LHC fixed at different distances from the beam core (more than $7\sigma$). TCSG.6R7.B1, placed at $7\sigma$, will be the main absorber. A high performance of the collimation system is expected to be achieved through the application of the channeling effect.

For the particles tracking in an accelerator a special routine was developed. It takes into account both betatron and synchrotron oscillations. The latter are simulated solving the differential equation system \cite{Kolom} describing the particle energy and synchrotron phase evolution with respect to RF cavities. The ionization energy losses as well as the energy losses due to quasielastic scattering in a crystal \cite{STRUCT} are also simulated. All the necessary parameters for the LHC were taken from the open sources \cite{LHCparam,LHCparamcoll}. The accelerator routine checks sequentially the possible collisions of particles with all the collimators. The tune shift due to the space charge in interaction points \cite{spacecharge} as well as the beam chromaticity \cite{Chrom} are also simulated. An option of a modified Monte Carlo method \cite{Sobol} is included. This method ascribes a unit weight to any particle. This weight decreases depending on the probability of the nuclear interaction in the crystal. The obtained weight decrement averaged over all the trajectories will give the percentage of the lost particles. The accuracy of this method will be higher than for the usual Monte-Carlo.

The distributions of the proton transverse horizontal coordinate at the absorber TCSG.6R7.B1 deflected by the (110) crystal at the channeling orientation is shown in Fig. 8. These distributions are similar to the angular distributions in Fig. 3 being obtained at the same conditions with the statistics of $10^{7}$ particles. Note that Fig. 3 describes the single-pass deflection while Fig. 8 concerns the multi-pass case which allows one to distinguish all the important effects such as channeling, dechanneling, volume reflection and strong single scattering by nuclei. Note that the multi-pass regime also allows one to observe the dechanneling peaks at the LHC.

\begin{figure}[t]
\hspace{-0.0cm} \vspace{0cm} {\includegraphics[width=10cm]{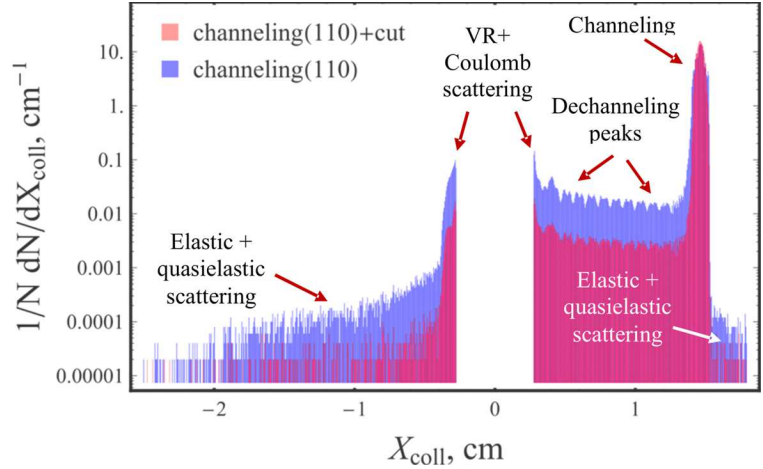}} \vspace{0cm}
 \caption {The transverse horizontal coordinate distribution at the absorber TCSG.6R7.B1 of 7 TeV protons after the deflection by the 4 mm crystal with and without the cut and bent by 50 $\mu$rad along the (110) planes at the channeling orientation for $10^{7}$ particles.}\label{F8}
\end{figure}

The channeling efficiency may be considerably improved with a crystal with the cut (Fig. 9a) \cite{cut,amlayer}. This technique allows one to considerably increase the efficiency of particle capture in the channeling mode by their focusing in the cut.  If the initial angular divergence is small enough (at least 2-3 times less than the Lindhard angle) we can find a universal condition for different particles to approach in the center of the channel after the crystal cut passage. Such particles will trap in the regime of stable channeling motion far from the crystal planes \cite{cut,amlayer}.

\begin{figure}[t]
\hspace{-0.0cm} \vspace{0cm} {\includegraphics[width=5cm]{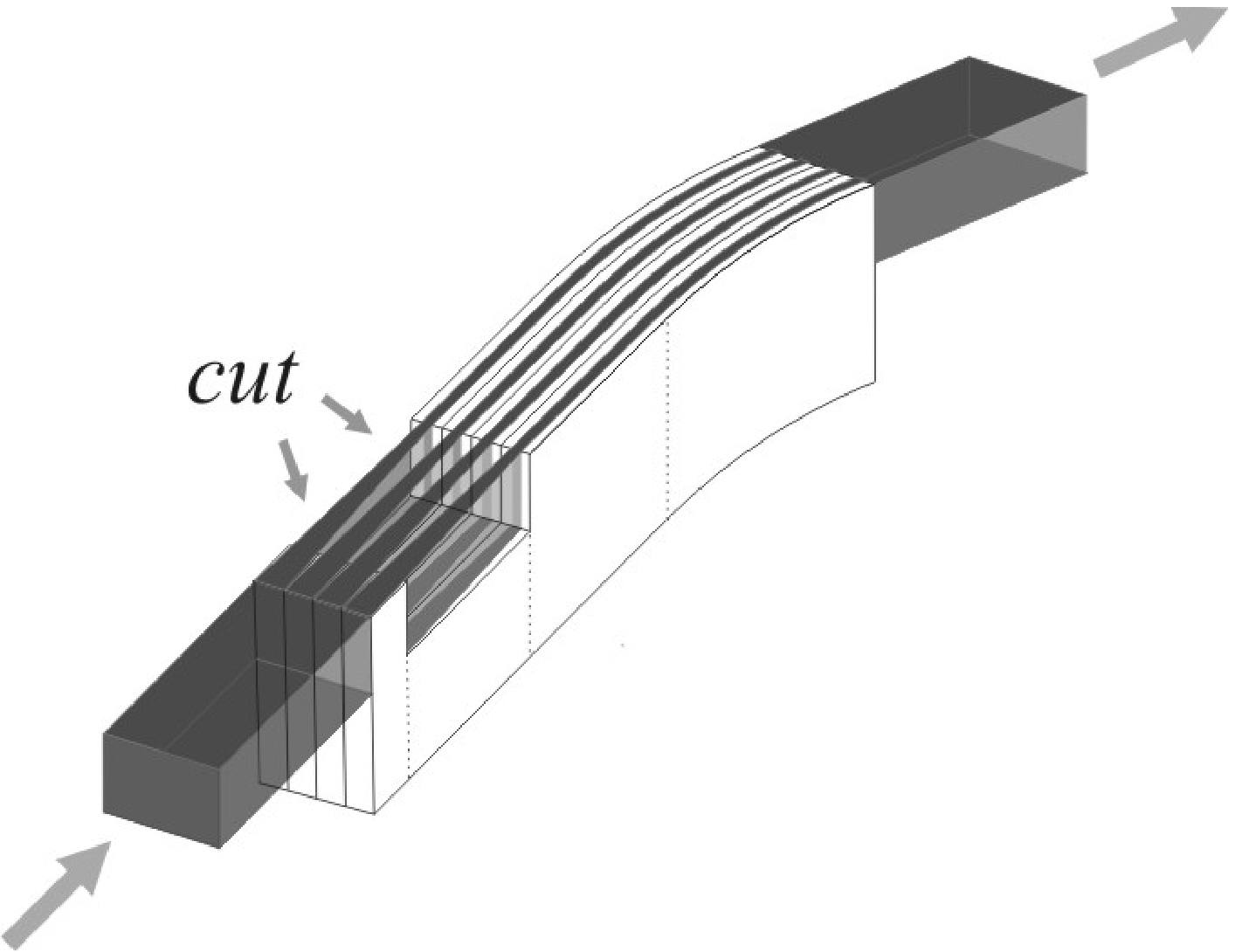}} \vspace{0cm} {\includegraphics[width=5cm]{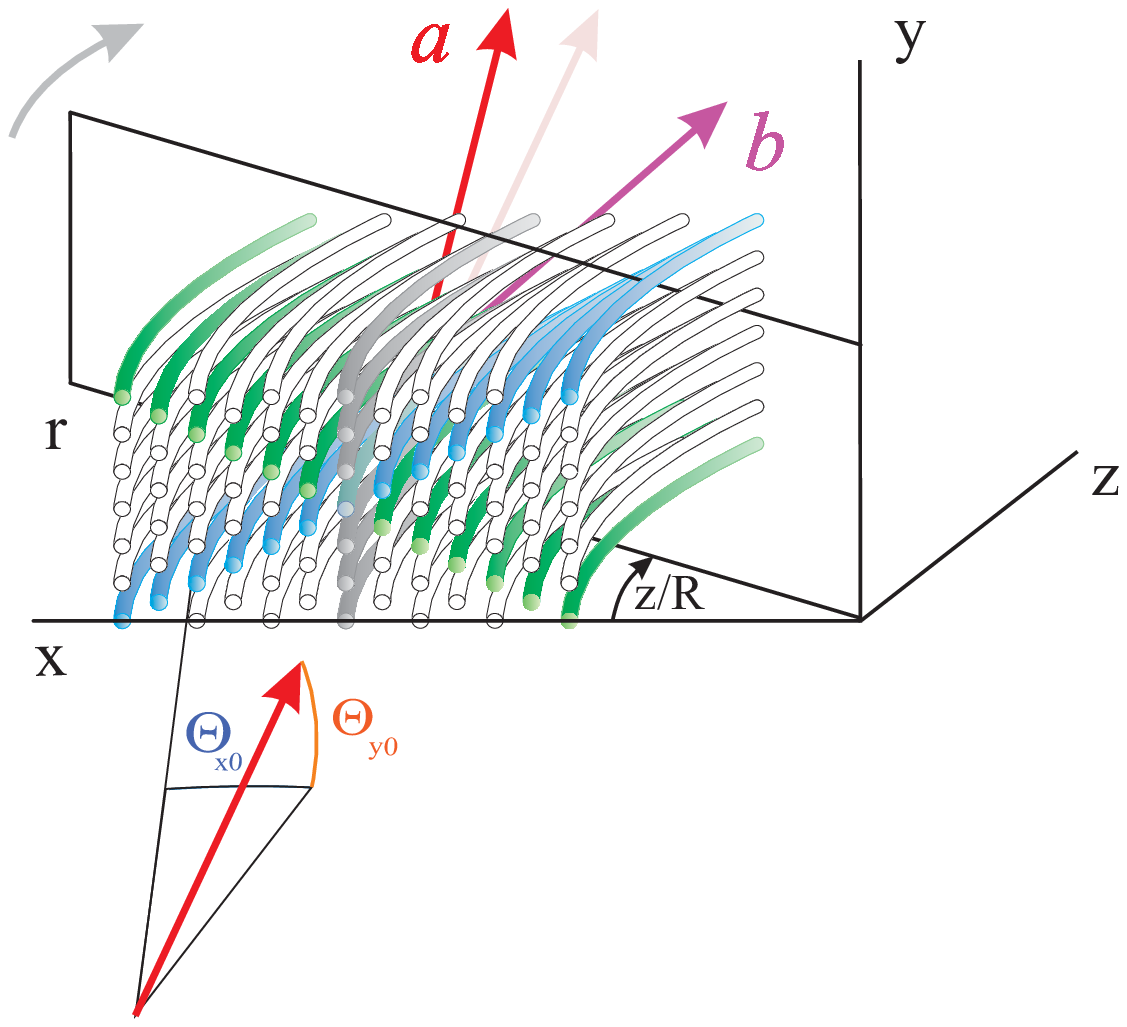}} \vspace{0cm}
 \caption {Bent crystal with the cut (left) and the effect of multiple volume reflection in one bent crystal (right).}\label{F9}
\end{figure}

The simulations of the LHC crystal-based collimation system with application of the crystal cut technique were conducted under the same conditions as the previous ones (Fig. 8). These results demonstrate a high performance of this method, in particular a order of value reduction of the quantity of dechanneled particles as well as of nuclear scattering events. Channeling single-pass deflection efficiency exceeds $98 \%$ which is almost $20 \%$ higher than for the crystal without the cut. It is important to add that dechanneling peaks are distinguishable in both cases.

In general, the crystal cut could become a smart solution to improve the performance of the crystal-assisted collimation. The most difficult is to provide very exact crystal alignment with the resolution less than 0.5 $\mu$rad. A bit lower accuracy is required for usual channeling which is also hard to achieve.

However, multiple volume reflection in one bent crystal (MVROC) \cite{MVROC,MVROCTGM,MVROCTS} providing high enough deflection angle doesn't require so precise crystal alignment. The idea of this effect consists in volume reflections from the skew crystal planes. While the vertical deflections substantially compensate each other the horizontal ones will be summarized (see Fig. 9 \cite{MVROC,MVROCTGM,MVROCTS}). The corresponding beam angular distribution behind the crystal is presented in Fig. 10. The crystal horizontal and vertical orientations were chosen correspondingly equal to $\theta_{Xcr}=-30$ $\mu$rad and $\theta_{Ycr}=-12$ $\mu$rad according to \cite{MVROCTS}. To compare the performance of the collimation system for all the cases the crystal length was chosen to be equal to 4 mm, i.e. the same like for channeling. Its bending angle was set equal to 60 $\mu$rad being twice higher than the reflecting particle incident angle w.r.t. to the crystal planes \cite{MVROCTS}. The crystal is bent along $\langle111\rangle$ axis.

Most of particles are deflected on approximately 10 $\mu$rad which is almost enough to get into the absorber at 7 TeV. However, it is possible to use advantages of both channeling and multiple volume reflection. Indeed, for a certain considered orientations of the crystal the capture into the channeling regime by a skew plane with inclination angle $\alpha_{pl}$ becomes possible \cite{MVROCTS} when:

\begin{equation}
\label{6}
\theta_{Ycr}=\theta_{Xcr}\tan \alpha_{pl},
\end{equation}
where $\alpha_{pl}$ is the angle of a chosen skew plane (($10\overline{1}$) plane in our case). The deflection angle distribution is illustrated in Fig. 10. One can see that most of particles will be deflected in the channeling mode by the angle sufficient for their absorbtion. Otherwise they will be reflected by means of multiple volume reflection in one bent crystal also providing a high angular kick.

\begin{figure}[b]
\hspace{-0.0cm} \vspace{0cm} {\includegraphics[width=10cm]{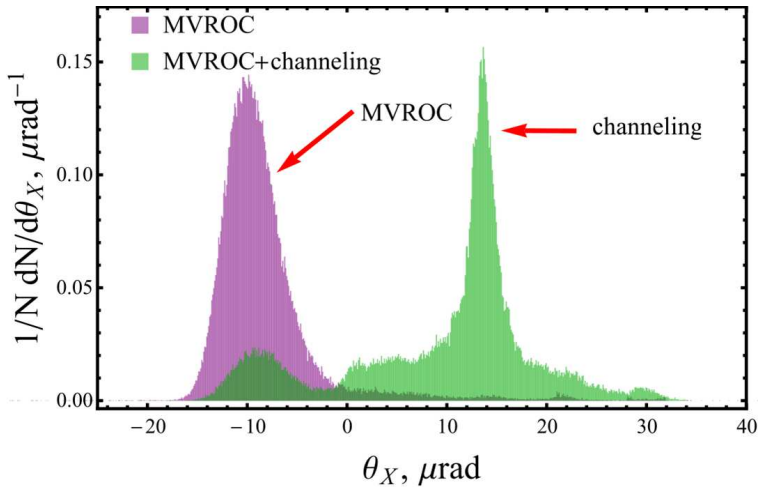}} \vspace{0cm}
 \caption {The angular distribution of 7 TeV protons behind the $\langle111\rangle$ bent crystal of 4 mm length and 60 $\mu$rad bending angle; horizontal alignment $\theta_{Xcr}=-30$ $\mu$rad, vertical alignment $\theta_{Ycr}=-12$ $\mu$rad for the MVROC and $\theta_{Ycr}=-17.3$ $\mu$rad for the combination of MVROC and channeling, $10^{6}$ particles.}\label{F10}
\end{figure}

The collimation efficiency dependence on the crystal alignment is shown in Fig. 11. We define the collimation efficiency as the beam fraction which wasn't lost inelastically in the crystal. The simulations were conducted for $10^{5}$--$10^{6}$ trajectories for each point and higher for the most important points.

Thus, the most effective is the channeling effect facilitated by the cut. The maximum efficiency is 99.95\% in this case. However, as it was mentioned above this method requires quite high angular resolution of 0.5 $\mu$rad which is close to the present technological limit. Some lower resolution is necessary for the usage of pure channeling in the (110) crystal providing the maximal efficiency of $99.66 \%$. The maximal performance of the (111) crystal is a bit worse.

At more practical resolution the combination of multiple volume reflection in one bent crystal and channeling becomes more effective. The efficiency of this method exceeds 99.5\% in this case. However, the collimation efficiency of the MVROC combined with channeling remains relatively high even in the case of a crystal misalignment. The level of the collimation efficiency for the pure MVROC is 98.8\% which is much higher than that of the usual volume reflection. The angular acceptance of the MVROC exceeds that of channeling by an order of value.

\begin{figure}[b]
\hspace{-0.0cm} \vspace{0cm} {\includegraphics[width=10cm]{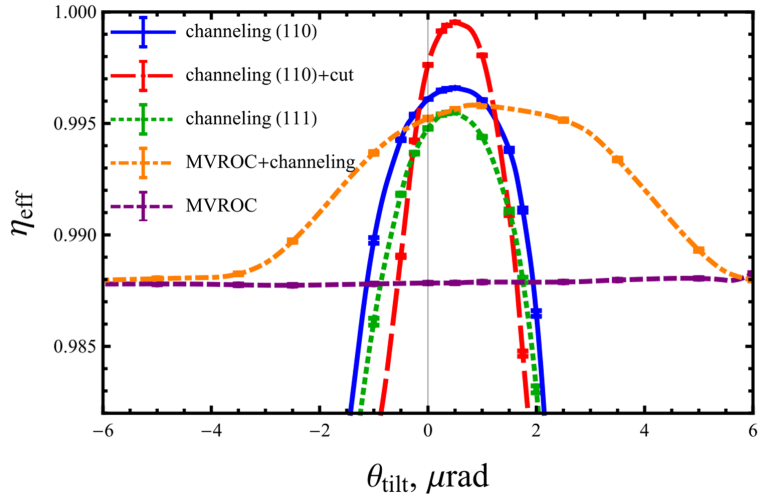}} \vspace{0cm}
 \caption {The dependence of the collimation efficiency on the crystal alignment for application of the channeling (in (110) and (111) planes), channeling in a crystal with the cut, MVROC and combined action of the MVROC and channeling effects.}\label{F11}
\end{figure}

Thus, the most efficient method for the crystal-based collimation at the LHC is the application of the crystal with the cut. However, if the angular resolution of the crystal alignment is worse than 0.5 $\mu$rad, the most efficient technique will be the combination of MVROC with channeling.

\section{Conclusions}

We implemented the charged particles trajectory simulation technique in the CRYSTAL software. This made possible both to reproduce of the H8 experiment at the CERN SPS and to predict the new effects. In particular, we predict that the dechanneling peaks representing planar channeling oscillations can be observed at high energies, specifically at 7 TeV in the LHC crystal-based collimation experiment. We found a condition of their appearance at the lower energy, especially at that of the SPS. The effect of ionization losses excess over the amorphous level for channeled particles with a critical amplitude was also revealed in CRYSTAL simulations.

The crystal cut may increase the collimation efficiency at the LHC up to 99.95\% because almost all the halo protons will be deflected onto secondary collimators in the channeling mode. However, requiring almost perfect crystal alignment the performance of such a technique may be considerably disturbed by a small crystal misorientation. In this case the combination of multiple volume reflection in one bent crystal and channeling in skew crystal planes becomes the most efficient for the LHC collimation.

Thus, the LHC crystal-based collimation system may be considerably improved by an application of either a bent crystal with the cut or combined action of MVROC and channeling.

\section{Acknowledgements}

We are grateful to Professors V.G. Baryshevsky, V. Guidi and I.D. Feranchuk for helpful advices and interesting discussions. We thank also the UA9 collaboration for useful discussions concerning the H8 experiment at the CERN SPS.



\begin{thebibliography}{21}

\bibitem{quench} J.B. Jeanneret et al., LHC Project Report 44 (1996) 16 p.
\bibitem{Tevatron} N.V. Mokhov et al., Intern. J. of Mod. Phys. A \textbf{25}, Suppl. 1 (2010) 98–105.
\bibitem{LUA9} D. Mirarchi et al., in: Proc. of IPAC, Dresden, Germany, June 15-20, 2014, MOPRI110 (882-886).
\bibitem{CrysColl} V.M. Biryukov et al., Nucl. Instr. Meth. in Phys. Res. B \textbf{234} (2005) 23-30.
\bibitem{cut} V.V. Tikhomirov, JINST \textbf{2} (2007) P08006 (1-11).
\bibitem{amlayer} V. Guidi, A. Mazzolari and V.V. Tikhomirov, J. Phys. D: Appl. Phys. \textbf{42} (2009) 165301 (1-6).
\bibitem{LHCExtr} E. Uggerh${\o}$j, U.I. Uggerh${\o}$j, Nucl. Instr. Meth. in Phys. Res. B \textbf{234} (2005) 31-39.
\bibitem{FCCExtr} A. Kovalenko in: Abstr. of  6th International Conference - Channeling 2014 Charged \& Neutral Particles Channeling Phenomena, Capri (Naples), Italy, October 5-10, 2014.
\bibitem{MVROC} V.V. Tikhomirov, Phys. Lett. B 655 (2007) 217-222.
\bibitem{MVROCTGM} V. Guidi, A. Mazzolari, and V.V. Tikhomirov, J. Appl. Phys. \textbf{107} (2010) 114908 (1-8).
\bibitem{MVROCTS} V.V. Tikhomirov, A.I. Sytov, Nucl. Instr. Meth. in Phys. Res. B \textbf{309} (2013) 109-114.
\bibitem{H8} R. Rossi, H8 Single Pass Test Data Analysis, http://lhc-collimation-upgrade-spec.web.cern.ch/LHC-Collimation-Upgrade-Spec/H8\_input.php\#cheff.
\bibitem{RungeCutta} G. I. Marchuk, Methods of Numerical Mathematics, Springer-Verlag, 1975, 316 p.
\bibitem{Tikh1} V. G. Baryshevskii, V. V. Tikhomirov, Zh. Eksp. Teor. Fiz. \textbf{90} (1986) 1116-1123.
\bibitem{Tikh2} V. V. Tikhomirov, Nucl. Instr. Meth. in Phys. Res. B \textbf{36} (1989) 282-285.
\bibitem{MokhovSc} A.I. Drozhdin et al., in: Proc. of PAC, Portland, Oregon, USA, May 12-16, 2003, pp. 1733-1735.
\bibitem{Biryukov} V.M.Biryukov, Y.A.Chesnokov, V.I.Kotov, Crystal Channeling and Its Application at High-Energy Accelerators, Springer-Verlag, 1997, 219 p.
\bibitem{DoyleTerner} S.L. Dudarev et al., Surface Science \textbf{330} (1995) 86-100.
\bibitem{H8no} R. Rossi et al., in: Presentation of Collimation Upgrade Meeting, August 22, 2014, http://lhc-collimation-upgrade-spec.web.cern.ch/LHC-Collimation-Upgrade-Spec/Files/meetings/44/colUSM\_22\_08\_14\_RR.pdf
\bibitem{H8line} M. Pesaresi et al., JINST \textbf{6} (2011) P04006.
\bibitem{miscut} V.V. Tikhomirov, A.I. Sytov, Problems of Atomic Science and Technology (Kharkov, Ukraine) \textbf{1}(2012)88-92; e-Print: arXiv:1109.5051.
\bibitem{LUA9old} D. Mirarchi et al., in: Proc. of IPAC, Shanghai, China, May 12-17, 2013, MOPRI035 (966-968).
\bibitem{Gemmel} D.S. Gemmel, Rev. Mod. Phys. \textbf{46} (1974) 129-227.
\bibitem{Feldman} L.C. Feldman, J.W. Mayer, S.T.A. Picraux, Materials Analysis by Ion Channeling: Submicron Crystallography, Academic Press (London), 1982.
\bibitem{Ba71} J. H. Barrett, Phys. Rev. B \textbf{3} (1971) 1527.
\bibitem{PCO0} F. Abel et al., Phys. Rev. B \textbf{12} N11 (1975) 4617-4627.
\bibitem{PCO1} M.B.H. Breese et al., Phys. Rev. B \textbf{53} N13 (1996) 8267-9276.
\bibitem{PCO2} M.B.H. Breese et al., Phys. Rev. Lett. \textbf{92} N13 (2004) 045503.
\bibitem{PDG} J. Beringer et al., Particle Data Group, Phys. Rev. D \textbf{86} (2012) 010001.
\bibitem{Kolom} A.A. Kolomensky, A.N. Lebedev, Circular accelerator theory, Moscow, 1962, 352 p.
\bibitem{STRUCT} A.I. Drozhdin et al. "STRUCT Program User's Reference.Manual", http://www-ap.fnal.gov/?drozhdin/.
\bibitem{LHCparam} LHC Optics Web Home http://lhc-optics.web.cern.ch/lhc-optics/www/.
\bibitem{LHCparamcoll} D. Mirarchi, Crystal Collimation http://lhc-collimation-upgrade-spec.web.cern.ch/LHC-Collimation-Upgrade-Spec/Sim7TeV\_crystals.php.
\bibitem{spacecharge} O.S. Br$\ddot{u}$ning. 2nd Evian 2010 Workshop on LHC Beam Operation (2010) 85-89.
\bibitem{Chrom} R. J. Steinhagen. CERN, Geneva, Switzerland, (2009) 43 p., http://cds.cern.ch/record/1213281/files/p317.pdf?version=1.
\bibitem{Sobol} I.M. Sobol, The Monte Carlo method, Science, Moscow, 1968, 63 p. (in Russian).

\end{thebibliography}
\end{document}